\documentclass[preprintnumbers,amsmath,amssymbm,prd]{revtex4}
\usepackage{epsfig}
\usepackage{graphicx}
\usepackage{amssymb}

\begin{document}
\title{A no-short scalar hair theorem for spinning acoustic black holes in a photon-fluid model}
\author{Shahar Hod}
\affiliation{The Ruppin Academic Center, Emeq Hefer 40250, Israel}
\affiliation{ } \affiliation{The Hadassah Institute, Jerusalem
91010, Israel}
\date{\today}

\begin{abstract}
\ \ \ It has recently been revealed that spinning black holes of the photon-fluid model 
can support acoustic `clouds', stationary density fluctuations whose spatially regular radial eigenfunctions are determined
by the $(2+1)$-dimensional Klein-Gordon equation of an effective massive scalar field. 
Motivated by this intriguing observation, we use {\it analytical} techniques in order to prove a no-short hair theorem for 
the composed acoustic-black-hole-scalar-clouds configurations. 
In particular, it is proved that the effective lengths of the stationary bound-state 
co-rotating acoustic scalar clouds are bounded from below by the series 
of inequalities $r_{\text{hair}}>{{1+\sqrt{5}}\over{2}}\cdot r_{\text{H}}>r_{\text{null}}$,
where $r_{\text{H}}$ and $r_{\text{null}}$ are respectively the horizon radius of the supporting black hole and 
the radius of the co-rotating null circular geodesic that characterizes the acoustic spinning black-hole spacetime.
\end{abstract}
\bigskip
\maketitle

%]

\section{Introduction}

Early mathematical studies of the Einstein-scalar field equations \cite{NS1,NS2,NS3,NS4}, which were motivated 
by the influential no-hair conjecture \cite{Whee,Car}, have revealed the physically interesting 
fact that asymptotically flat 
black holes with regular horizons cannot support in their exterior regions 
static matter configurations which are made of minimally coupled scalar fields.  

However, subsequent analyzes (see \cite{BizCol,Lavr,BizCham,Green,Stra,BiWa,EYMH,Volkov,BiCh,Lav1,Lav2,Bizw} and references therein) 
of the Einstein-matter field equations have explicitly demonstrated that
black-hole spacetimes may not be as simple as 
suggested by the original no-hair conjecture \cite{Whee,Car}. In particular, it is by now well 
established in the physics literature \cite{BizCol,Lavr,BizCham,Green,Stra,BiWa,EYMH,Volkov,BiCh,Lav1,Lav2,Bizw} 
that spherically symmetric asymptotically flat black holes {\it can} support various types of hairy matter configurations, 
static fields which are well behaved on and outside the black-hole horizon.  

In addition, it has been proved analytically \cite{Hodrc} that the superradiant 
scattering phenomenon of bosonic fields in spinning black-hole spacetimes \cite{Zel,PressTeusup} allows non-static 
Kerr black holes to support stationary 
bound-state matter configurations 
which are made of minimally coupled linearized massive scalar fields. 
These externally supported scalar field configurations, which co-rotate with the central spinning black hole, 
have received the nickname `scalar clouds' in the linearized regime \cite{Hodrc,HerR}. 
Interestingly, using sophisticated numerical techniques, the existence of genuine hairy (scalarized) spinning 
black-hole solutions of the non-linearly coupled Einstein-scalar field equations has been explicitly demonstrated in \cite{HerR}.

Interestingly, the stationary co-rotating externally supported bosonic field configurations are characterized by 
proper frequencies which are in resonance 
with the horizon angular velocity of the central supporting black hole \cite{Hodrc,HerR,Noteunits,Notemmmm},
\begin{equation}\label{Eq1}
\omega=m\Omega_{\text{H}}\  .
\end{equation}
In addition, the proper frequencies of the supported bound-state field configurations 
are bounded from above by the proper mass of the supported scalar field \cite{Notedim}:
\begin{equation}\label{Eq2}
\omega^2<\mu^2\  .
\end{equation}

Given the physically intriguing fact that hairy black-hole solutions of the Einstein-matter field equations do exist, one 
may raise the following physically interesting question: How short can a black-hole hair be?

For static spherically symmetric hairy black-hole spacetimes, the answer to this question has been provided in \cite{Hodshort}, where 
it was proved that the effective lengths of spatially regular hairy matter configurations whose energy-momentum trace is non-positive  
must extend beyond the innermost null circular geodesics of the corresponding curved black-hole spacetimes:
\begin{equation}\label{Eq3}
r_{\text{hair}}>r_{\text{null}}\  .
\end{equation}
As explicitly proved in \cite{HodshortKerr}, the effective lengths of the
co-rotating non-spherically symmetric scalar cloudy configurations of the spinning Kerr spacetime \cite{Hodrc,HerR} 
also conform to the lower bound (\ref{Eq3}). 

Interestingly, it is well established that fluid systems share many features with curved 
black-hole spacetimes (see \cite{Un,Cim,Mar1,PF1,PF2,NR1,NR2,NR3,NR4,NRR1,NRR2,NRR3,NRR4,NRRR1,NRRR2,Hodpf21} 
and references therein). 
In particular, it has recently been proved in the physically important work \cite{Cim} 
that acoustic black holes of the $(2+1)$-dimensional rotating photon-fluid system
can support stationary bound-state density fluctuations (acoustic scalar `clouds') whose spatio-temporal behavior 
in the black-hole spacetime 
is governed by the linearized Klein-Gordon equation of an effective massive scalar field. 

As nicely emphasized in \cite{Cim}, the co-rotating acoustic scalar clouds of the photon-fluid model, like the more familiar  
scalar hairy configurations of the Kerr black-hole spacetime \cite{Hodrc,HerR}, owe their 
existence to the physically intriguing phenomenon of superradiant scattering of co-rotating bosonic field modes 
in the spinning physical system. In particular, the $(2+1)$-dimensional stationary acoustic clouds revealed in \cite{Cim} 
are characterized by the same resonance condition [see Eq. (\ref{Eq1})] as the Kerr scalar clouds \cite{NoteOmgh}. 

The main goal of the present paper is to analyze the spatial functional behavior 
of the stationary bound-state acoustic scalar field configurations (linearized scalar clouds) that are supported by 
the effective spinning black-hole spacetime of the photon-fluid model \cite{Cim}. 
In particular, motivated by the existence of the lower bound (\ref{Eq3}) on the effective lengths of 
hairy matter configurations in the black-hole spacetime solutions of the Einstein field equations, we shall use 
analytical techniques in order to derive an analogous generic lower bound on the effective lengths of the 
composed acoustic-black-hole-scalar-field cloudy configurations of the physically interesting photon-fluid model.

\section{Description of the system}

The spinning $(2+1)$-dimensional acoustic black-hole spacetime of the photon-fluid model is characterized 
by the curved line element \cite{Cim}
\begin{equation}\label{Eq4}
ds^2=-\Big(1-{{r_{\text{H}}}\over{r}}-{{\Omega^2_{\text{H}}r^4_{\text{H}}}\over{r^2}}\Big)dt^2+
\Big(1-{{r_{\text{H}}}\over{r}}\Big)^{-1}dr^2-
2\Omega_{\text{H}}r^2_{\text{H}}d\theta dt+r^2 d\theta^2\  ,
\end{equation}
where $\{r,\theta\}$ are the familiar polar coordinates in a two-dimensional plane and the 
physical parameters $\{r_{\text{H}},\Omega_{\text{H}}\}$ are respectively the horizon radius \cite{Notehor}
and the angular velocity of the spinning acoustic horizon. 
The acoustic black-hole spacetime (\ref{Eq4}), like the spinning Kerr black-hole spacetime, possesses 
an ergoregion whose outer radial location \cite{Cim}
\begin{equation}\label{Eq5}
r_{\text{E}}={{1}\over{2}}r_{\text{H}}\Big(1+\sqrt{1+4\Omega^2_{\text{H}}r^2_{\text{H}}}\Big)\ 
\end{equation}
is determined by the root of the metric function $g_{tt}$. 

As explicitly shown in \cite{Cim,Mar1}, long-wavelength excitations (phonons) of the photon-fluid system
behave as effective massive scalar fields that propagate in the acoustic curved spacetime (\ref{Eq4}). 
In particular, given a linearized acoustic density fluctuation \cite{Notemmm}
\begin{equation}\label{Eq6}
\rho(t,r,\theta)={{\psi(r)}\over{\sqrt{r}}}e^{im\theta-i\Omega t}\  
\end{equation} 
of the photon-fluid model, it has been proved that its spatio-temporal behavior is determined by the $(2+1)$-dimensional 
Klein-Gordon differential equation \cite{Cim,Mar1}
\begin{equation}\label{Eq7}
\Big[\Delta{{d}\over{dr}}\Big(\Delta{{d}\over{dr}}\Big)-V(r;\Omega)\Big]\psi(r)=0\ \ \ \ ; \ \ \ \ 
\Delta\equiv 1-{{r_{\text{H}}}\over{r}}\  .
\end{equation}  
The radial potential \cite{Cim}
\begin{equation}\label{Eq8}
V(r;\Omega)=-\Big(\Omega-{{m\Omega_{\text{H}}r^2_{\text{H}}}\over{r^2}}\Big)^2+
\Delta\Big(\Omega^2_0+{{m^2}\over{r^2}}+{{r_{\text{H}}}\over{2r^3}}-{{\Delta}\over{4r^2}}\Big)\  ,
\end{equation}
which determines the spatial behavior of the density fluctuations (\ref{Eq6}) 
in the acoustic curved spacetime (\ref{Eq4}), corresponds to 
an effective scalar field $\psi$ of mass $\Omega_0$ \cite{Notemass0,Noteflm}.

In the present paper we shall use analytical techniques in order to analyze the spatial behavior of 
the composed acoustic-black-hole-stationary-linearized-massive-scalar-field configurations of the 
photon-fluid system. 
The stationary bound-state scalar clouds of the spinning acoustic spacetime (\ref{Eq4}) are characterized by the 
resonant frequency \cite{Noteana1}
\begin{equation}\label{Eq9}
\Omega=m\Omega_{\text{H}}\  .
\end{equation}
In addition, the scalar eigenfunctions of the supported acoustic clouds are assumed to be regular 
at the acoustic black-hole horizon \cite{Cim}: 
\begin{equation}\label{Eq10}
\psi(r=r_{\text{H}})<\infty\  .
\end{equation}
The bound-state acoustic scalar eigenfunctions are also assumed to be normalizable (decay exponentially) 
at spatial infinity \cite{Cim}:
\begin{equation}\label{Eq11}
\psi(r\to\infty)\sim e^{-{\sqrt{\Omega^2_0-\Omega^2}r}}\ 
\end{equation}
for \cite{Noteana2} 
\begin{equation}\label{Eq12}
\Omega^2<\Omega^2_0\  .
\end{equation}

As demonstrated numerically in \cite{Cim} and proved analytically in \cite{Hodpf21}, 
the stationary bound-state acoustic-black-hole-massive-scalar-field cloudy configurations of 
the photon-fluid model, which respect the boundary conditions (\ref{Eq10}) and (\ref{Eq11}), 
are characterized by the dimensionless regime of existence
\begin{equation}\label{Eq13}
{{\Omega_0}\over{m\Omega_{\text{H}}}}\in\Big(1,\sqrt{{{32}\over{27}}}\Big)\  .
\end{equation}

In the next section we shall reveal, using analytical techniques, the existence of a generic lower bound 
on the effective radial lengths of the supported co-rotating acoustic scalar clouds of the photon-fluid model. 

\section{Lower bound on the effective radial lengths of the stationary bound-state acoustic
scalar clouds of the photon-fluid model}

In the present section we shall explore the spatial functional behavior of the 
scalar eigenfunctions $\psi(r;r_{\text{H}},\Omega_{\text{H}},\Omega_0,m)$ which
characterize the the linearized massive scalar field configurations (stationary scalar clouds) that are supported by the 
$(2+1)$-dimensional acoustic black-hole spacetime (\ref{Eq4}) of the photon-fluid model \cite{Cim}. 
In particular, we shall explicitly prove that the stationary bound-state acoustic scalar clouds cannot be arbitrarily compact. 

To this end, we shall first prove that the bound-state scalar clouds of the photon-fluid model are 
characterized by a {\it non}-monotonic radial eigenfunction $\psi(r)$. 
We shall then derive, using the explicit functional behavior of the effective radial potential (\ref{Eq8}), 
a generic (parameter-independent) lower bound [see Eq. (\ref{Eq24}) below]
on the peak location $r_{\text{max}}$ of the radial scalar eigenfunctions that characterize the 
supported co-rotating acoustic scalar clouds of the photon-fluid model.

Before proceeding, we would like to emphasize that the interesting lower bound 
\begin{equation}\label{Eq14}
{{r_{\text{min}}}\over{r_{\text{H}}}}>{{3}\over{2(\Omega_{\text{H}}r_{\text{H}})^2}}\
\end{equation}
on the radial location of the minimum $r=r_{\text{min}}$ of the effective potential (\ref{Eq8}) \cite{Noteminv}
has been derived in the physically important work \cite{Cim}. The lower bound (\ref{Eq14}) of \cite{Cim} 
nicely demonstrates the physically important fact that, in the slow rotation $\Omega_{\text{H}}\to0$ limit of the 
central supporting acoustic black holes, the scalar clouds are effectively located far away from the central black hole. 
However, it should be realized that the interesting bound (\ref{Eq14}), which is based on 
the asymptotic large-$r$ expansion of the effective radial potential (\ref{Eq8}) [see \cite{Cim} for details], 
is unable to describe the genuine near-horizon radial behavior of the stationary scalar clouds in the 
regime $\Omega_{\text{H}}r_{\text{H}}\gg1$ of rapidly-spinning central supporting acoustic black holes \cite{Notemoi}. 
In particular, one finds that the right-hand-side of (\ref{Eq14}) is less than $1$ 
%${{r_{\text{min}}}<{r_{\text{H}}}}$ 
for $\Omega_{\text{H}}r_{\text{H}}\gtrsim1$, thus suggesting that the bound (\ref{Eq14}) works well 
for slowly rotating black holes (for which $r_{\text{min}}\gg r_{\text{H}}$) but breaks down for 
rapidly-spinning acoustic black holes.

In the present section we shall use the exact functional form of the composed acoustic-black-hole-massive-scalar-field binding  
potential (\ref{Eq8}) in order to derive an alternative lower bound on the peak location $r_{\text{max}}$ of the 
radial eigenfunctions $\psi(r;r_{\text{H}},\Omega_{\text{H}},\Omega_0,m)$ that characterize the 
bound-state linearized scalar clouds of the photon-fluid model. In particular, we shall 
explicitly prove below that, for{\it all} values of the dimensionless rotation parameter $\Omega_{\text{H}}r_{\text{H}}$ 
of the central supporting acoustic black hole, the peak location $r=r_{\text{max}}$ of the acoustic 
scalar configurations cannot be located arbitrarily close to the black-hole horizon. 

The radial functional behavior of the stationary bound-state cloudy field configurations of the 
photon-fluid system is determined by the 
ordinary differential equation (\ref{Eq7}) with the effective binding potential [see Eqs. (\ref{Eq8}) and (\ref{Eq9})] 
\begin{equation}\label{Eq15}
V(r)=-(m\Omega_{\text{H}})^2\cdot\Big(1-{{r^2_{\text{H}}}\over{r^2}}\Big)^2+
\Delta\Big(\Omega^2_0+{{m^2}\over{r^2}}+{{r_{\text{H}}}\over{2r^3}}-{{\Delta}\over{4r^2}}\Big)\  .
\end{equation}
We first point out that in the near-horizon region,
\begin{equation}\label{Eq16}
x\equiv {{r-r_{\text{H}}}\over{r_{\text{H}}}}\ll1\  ,
\end{equation}
the effective radial potential (\ref{Eq15}) of the 
composed acoustic-black-hole-stationary-bound-state-massive-scalar-field 
configurations is characterized by the functional behavior
\begin{equation}\label{Eq17}
V(x\ll1)=\Big(\Omega^2_0+{{m^2+{1\over2}}\over{r^2_{\text{H}}}}\Big)\cdot x+O(x^2/r^2_{\text{H}})\  ,
\end{equation}
which implies 
\begin{equation}\label{Eq18}
V(x)>0\ \ \ \ \text{for}\ \ \ \ 0<x\ll1\  .
\end{equation}

We shall now consider two mathematically distinct cases for the possible near-horizon functional behaviors
of the scalar eigenfunction $\psi(r)$:

Case (i): If $\psi(r=r_{\text{H}})=0$, then the asymptotic boundary condition (\ref{Eq11}), 
which characterizes the radial behavior of the bound-state cloudy scalar configurations at spatial infinity, implies 
that the scalar eigenfunction $\psi(r)$ must have an extremum point $r=r_{\text{max}}$ \cite{Notepwlg} in the exterior region of the 
effective black-hole spacetime. 

Case (ii): If $\psi(r\to r_{\text{H}})\neq0$ and $[d\psi(r)/dr]_{r=r_{\text{H}}}\neq0$ \cite{Notepdr0}, 
then one deduces from the 
radial differential equation (\ref{Eq7}) and the near-horizon functional behavior (\ref{Eq17}) of the 
effective radial potential that $\psi(r)\cdot d\psi(r)/dr>0$ for $r\to r_{\text{H}}$. 
This observation together with the asymptotic boundary condition (\ref{Eq11}) imply again that the 
scalar eigenfunction $\psi(r)$, which characterizes the spatial behavior of the acoustic scalar clouds, must have an extremum point $r=r_{\text{max}}$ in the exterior region of the spinning acoustic black-hole spacetime. 

We therefore conclude that the stationary bound-state scalar clouds of the photon-fluid model are characterized by 
non-monotonic radial eigenfunctions. I particular, the acoustic scalar eigenfunction $\psi(r)$ is characterized by the 
presence of an extremum radial point $r=r_{\text{max}}$ in the exterior region of the acoustic black-hole spacetime 
with the properties
\begin{equation}\label{Eq19}
\Big\{\psi\neq0\ \ \ ; \ \ \ {{d\psi}\over{dr}}=0\ \ \ ; \ \ \ \psi{{d^2\psi}\over{dr^2}}<0\Big\}\ \ \ \ \text{for}\ \ \ \ r=r_{\text{max}}\  .
\end{equation}

Substituting the characteristic functional relations (\ref{Eq19}) into the radial differential equation (\ref{Eq7}), one 
finds the simple relation
\begin{equation}\label{Eq20}
V(r=r_{\text{max}})<0\  .
\end{equation}
Taking cognizance of Eqs. (\ref{Eq15}) and (\ref{Eq20}), one finds the characteristic series of 
inequalities \cite{Notemge1} 
\begin{equation}\label{Eq21}
%\Omega^2_0\cdot\Big(1-{{r^2_{\text{H}}}\over{r^2}}\Big)^2>
(m\Omega_{\text{H}})^2\cdot\Big(1-{{r^2_{\text{H}}}\over{r^2}}\Big)^2>
\Delta\Big(\Omega^2_0+{{m^2}\over{r^2}}+{{r_{\text{H}}}\over{2r^3}}-{{\Delta}\over{4r^2}}\Big)>
\Delta\cdot\Omega^2_0\ \ \ \ \text{for}\ \ \ \ r=r_{\text{max}}\  ,
\end{equation}
which implies [see Eq. (\ref{Eq7})]
\begin{equation}\label{Eq22}
\Big(1-{{r_{\text{H}}}\over{r_{\text{max}}}}\Big)\Big(1+{{r_{\text{H}}}\over{r_{\text{max}}}}\Big)^2>
\Big({{\Omega_0}\over{m\Omega_{\text{H}}}}\Big)^2\  .
%%\ \ \ \ \text{for}\ \ \ \ r=r_{\text{max}}\  .
%{{\Big(1-{{r^2_{\text{H}}}\over{r^2}}\Big)^2}\over{1-{{r_{\text{H}}}\over{r}}}}>
%\Big({{\Omega_0}\over{m\Omega_{\text{H}}}}\Big)^2\ \ \ \ \text{for}\ \ \ \ r=r_{\text{max}}\  .
%%\Big(1-{{r^2_{\text{H}}}\over{r^2}}\Big)^2>1-{{r_{\text{H}}}\over{r}}\ \ \ \ \text{for}\ \ \ \ r=r_{\text{max}}\  .
\end{equation}

From the analytically derived cubic inequality (\ref{Eq22}) one obtains the dimensionless lower bound
\begin{equation}\label{Eq23}
{{r_{\text{max}}}\over{r_{\text{H}}}}>F\Big({{\Omega_0}\over{m\Omega_{\text{H}}}}\Big)\
\end{equation}
on the location $r=r_{\text{max}}$ of the radial peak of the acoustic scalar eigenfunctions, 
where the (mathematically cumbersome) dimensionless function $F=F\big({{\Omega_0}\over{m\Omega_{\text{H}}}}\big)$ 
is a monotonically increasing function in the regime of existence ${{\Omega_0}/{m\Omega_{\text{H}}}}\in(1,\sqrt{32/27})$ 
[see Eq. (\ref{Eq13})] of the 
composed acoustic-black-hole-stationary-bound-state-massive-scalar-field configurations. 
In particular, from (\ref{Eq22}) one directly finds that the function $F\big({{\Omega_0}\over{m\Omega_{\text{H}}}}\big)$ 
in the lower bound (\ref{Eq23}) increases
from $F\big({{\Omega_0}\over{m\Omega_{\text{H}}}}\to1^{+}\big)\to[(1+\sqrt{5})/2]^+$ to 
$F\big({{\Omega_0}\over{m\Omega_{\text{H}}}}\to{\sqrt{32/27}}^{-}\big)\to3^-$. 
We therefore find the generic (that is, rotation-independent) 
lower bound 
\begin{equation}\label{Eq24}
{{r_{\text{max}}}\over{r_{\text{H}}}}>{{1+\sqrt{5}}\over{2}}\
\end{equation}
on the effective radial lengths of the stationary bound-state acoustic scalar clouds which are supported 
by the spinning black-hole spacetime (\ref{Eq4}). 

It is interesting to emphasize the fact that the lower bound (\ref{Eq24}) on the 
effective lengths of the co-rotating acoustic scalar clouds is universal in the sense that it does not depend 
on the physical parameters (proper mass $\Omega_0$ and azimuthal harmonic index $m$) of the supported 
acoustic scalar field.

\section{Co-rotating acoustic scalar clouds and null circular geodesics}

In the present section we shall explicitly prove that the 
composed acoustic-black-hole-stationary-bound-state-linearized-massive-scalar-field configurations 
of the photon-fluid model, like the scalarized spinning black-hole solutions of the Einstein field equations, conform 
to the no-short hair relation (\ref{Eq3}). 
In particular, as we shall now show, the radial peak location $r_{\text{max}}$, which characterizes the non-monotonic 
eigenfunctions $\psi(r)$ of the bound-state acoustic scalar clouds, is located beyond the co-rotating null circular
geodesic of the effective spinning black-hole spacetime (\ref{Eq4}).

A remarkably economic way to determine the radial location of the co-rotating null circular geodesic of 
a curved black-hole spacetime has been revealed in \cite{Hodnull}. In particular, it has been proved in \cite{Hodnull,Yng} that the 
co-rotating null circular geodesic provides the fastest way, as measured by asymptotic observers, 
to circle the central black hole. Substituting $ds=dr=0$ and $d\theta=2\pi$ into the curved line element (\ref{Eq4}), one finds the 
functional expression 
\begin{equation}\label{Eq25}
{{T(r)}\over{r_{\text{H}}}}=2\pi\Omega_{\text{H}}r_{\text{H}}\cdot
{{\sqrt{1+{{r^2}\over{\Omega^2_{\text{H}}r^4_{\text{H}}}}\cdot\Big(1-{{r_{\text{H}}}\over{r}}-{{\Omega^2_{\text{H}}r^4_{\text{H}}}\over{r^2}}\Big)}-1}\over{1-{{r_{\text{H}}\over{r}}}-{{\Omega^2_{\text{H}}r^4_{\text{H}}}\over{r^2}}}}
\end{equation}
for the (radius-dependent) dimensionless orbital period of light-like test particles around the central black hole \cite{Notevcc}. 

As explicitly shown in \cite{Hodnull,Yng}, the co-rotating null circular geodesics of curved black-hole spacetimes are 
characterized by the relation
\begin{equation}\label{Eq26}
{{dT(r)}\over{dr}}=0\ \ \ \ \text{for}\ \ \ \ r=r_{\text{null}}\  .
\end{equation}
Substituting (\ref{Eq25}) into Eq. (\ref{Eq26}), one obtains the characteristic equation
\begin{equation}\label{Eq27}
{{r^2(r-r_{\text{H}})(2r-3r_{\text{H}})+r(5r_{\text{H}}-6r)\Omega^2_{\text{H}}r^4_{\text{H}}+
2\Omega_{\text{H}}r^3_{\text{H}}\sqrt{r(r-r_{\text{H}})}(r+2\Omega^2_{\text{H}}r^3_{\text{H}})}\over
{\Big(1-{{r_{\text{H}}}\over{r}}-{{\Omega^2_{\text{H}}r^4_{\text{H}}}\over{r^2}}\Big)^2}}=0\ \ \ \ \text{for}\ \ \ \ r=r_{\text{null}}\
\end{equation}
for the radial location $r=r_{\text{null}}$ of the co-rotating null circular geodesic of the 
acoustic spinning black-hole spacetime (\ref{Eq4}). 

From Eq. (\ref{Eq27}) one finds that the $\Omega_{\text{H}}r_{\text{H}}$-dependent radial location $r_{\text{null}}=r_{\text{null}}(\Omega_{\text{H}}r_{\text{H}})$ 
of the co-rotating null circular geodesic is restricted to the interval 
\begin{equation}\label{Eq28}
{{r_{\text{null}}}\over{r_{\text{H}}}}\in(1,{{3}\over{2}}]\  .
\end{equation}
In particular, from (\ref{Eq27}) one finds that $r_{\text{null}}(\Omega_{\text{H}}r_{\text{H}})$ is a monotonically decreasing function 
of the dimensionless black-hole rotation parameter $\Omega_{\text{H}}r_{\text{H}}$ with the simple asymptotic behaviors 
\begin{equation}\label{Eq29}
{{r_{\text{null}}}\over{r_{\text{H}}}}={3\over2}-{{2}\over{\sqrt{3}}}\cdot\Omega_{\text{H}}r_{\text{H}}+
O[(\Omega_{\text{H}}r_{\text{H}})^2]
\ \ \ \ \text{for}\ \ \ \ \Omega_{\text{H}}r_{\text{H}}\ll1\
\end{equation}
and 
\begin{equation}\label{Eq30}
{{r_{\text{null}}}\over{r_{\text{H}}}}=1+{{1}\over{16(\Omega_{\text{H}}r_{\text{H}})^2}}+
O[(\Omega_{\text{H}}r_{\text{H}})^{-3}]
\ \ \ \ \text{for}\ \ \ \ \Omega_{\text{H}}r_{\text{H}}\gg1\  .
\end{equation}

Taking cognizance of Eqs. (\ref{Eq24}) and (\ref{Eq28}), one concludes that the stationary
bound-state scalar clouds of the spinning acoustic black-hole spacetime (\ref{Eq4})
are characterized by the lower bound
\begin{equation}\label{Eq31}
r_{\text{max}}>r_{\text{null}}\  .
\end{equation}

\section{Summary}

A decade ago it has been proved that spinning Kerr black holes can support co-rotating scalar clouds, 
stationary bound-state linearized 
configurations of spatially regular massive scalar fields whose orbital frequencies are in resonance 
with the angular velocity $\Omega_{\text{H}}$ of the black-hole horizon \cite{Hodrc,HerR}. 
The bound-state scalar configurations are known to be characterized by the
no-short hair property \cite{Hodshort,HodshortKerr}, according to which their effective lengths 
extend beyond the null circular geodesics of the supporting black-hole spacetimes.
 
Intriguingly, it has recently been revealed in the physically important work \cite{Cim} 
that an analogous physical phenomenon occurs in 
a rotating photon-fluid model \cite{Cim}. 
In particular, it has been demonstrated \cite{Cim} that in the presence of vortex flows, the photon-fluid 
system may be described by an effective rotating acoustic black-hole spacetime [see Eq. (\ref{Eq4})] which, like the spinning Kerr black-hole spacetime, 
may support stationary linearized density fluctuations (acoustic scalar clouds) 
whose spatial behavior is governed by the Klein-Gordon equation of a $(2+1)$-dimensional 
scalar field with an effective proper mass $\Omega_0$. 

The main goal of the present paper was to analyze the spatial behavior 
of the co-rotating acoustic scalar clouds that are supported by 
the $(2+1)$-dimensional acoustic black hole (\ref{Eq4}) of the photon-fluid model. 
Interestingly, we have established the fact that the supported acoustic 
scalar configurations cannot be made
arbitrarily compact. 
In particular, using analytical techniques, we have derived a generic lower bound on the effective lengths of the 
bound-state acoustic-black-hole-scalar-field cloudy configurations of the photon-fluid model. 
This parameter-independent bound can be expressed in a remarkably compact way by the dimensionless series 
of inequalities [see Eqs. (\ref{Eq24}) and (\ref{Eq31})]
\begin{equation}\label{Eq32}
r_{\text{max}}>{{1+\sqrt{5}}\over{2}}\cdot r_{\text{H}}>r_{\text{null}}\  ,
\end{equation}
where $\{r_{\text{H}},r_{\text{null}}\}$ are respectively the horizon radius and 
the radius of the co-rotating null circular geodesic that characterize the supporting acoustic black-hole spacetime.

Finally, it is worth emphasizing the physically interesting fact that 
the analytically derived lower bound (\ref{Eq32}) on the effective lengths of the bound-state
acoustic scalar clouds of the photon-fluid model
is universal in the sense that it is valid for {\it all} possible sets $\{r_{\text{H}},\Omega_{\text{H}},\Omega_0,m\}$ 
of the physical parameters that characterize the supporting spinning acoustic black hole 
and the effective massive scalar fields.

\bigskip
\noindent
{\bf ACKNOWLEDGMENTS}
\bigskip

This research is supported by the Carmel Science Foundation. I thank
Yael Oren, Arbel M. Ongo, Ayelet B. Lata, and Alona B. Tea for
stimulating discussions.

%\newpage


\begin{thebibliography}{99}
\bibitem{Notemmmm} Here the integer $m$ is the azimuthal harmonic index that characterizes 
the supported scalar eigenfunction [see Eq. (\ref{Eq6}) below].
\bibitem{Noteunits} We use natural units in which $G=c=\hbar=1$.
\bibitem{Notedim} Note that the mass parameter $\mu$, which characterizes the scalar field, 
stands for $\mu/\hbar$. Hence, this physical parameter has the dimensions of (length)$^{-1}$.
\bibitem{NoteOmgh} For the co-rotating acoustic scalar clouds discussed in \cite{Cim}, the physical parameter $\Omega_{\text{H}}$ 
in the resonance condition 
is the horizon angular velocity of the central supporting $(2+1)$-dimensional acoustic black hole.
\bibitem{Notehor} The radius $r=r_{\text{H}}$ of the acoustic black-hole horizon 
is determined by the physical requirement $v_{\text{r}}=c_{\text{s}}$, where 
$\{v_{\text{r}},c_{\text{s}}\}$ are respectively the 
inward radial velocity of the fluid flow and the speed of sound in the fluid \cite{Cim}. 
We shall use natural units in which $c_{\text{s}}\equiv1$.
\bibitem{Notemmm} Note that the circular periodicity of the azimuthal eigenfunction $e^{im\theta}$ 
implies that the discrete field parameter $m$ is an integer.
\bibitem{Notemass0} As discussed in \cite{Cim}, the effective scalar mass $\Omega_0$ in the composed 
acoustic-black-hole-scalar-field curvature potential (\ref{Eq8}) is the rest
energy of the collective excitations (phonons) \cite{Cim}.
\bibitem{Noteflm} Note that the flat-space $r\to\infty$ limit of the effective potential (\ref{Eq8}) 
is given by the simple functional expression $V(r\to\infty)=-\Omega^2+\Omega^2_0$, and it 
therefore describes an effective scalar field of proper frequency $\Omega$ and an effective proper mass $\Omega_0$.
\bibitem{Noteminv} As emphasized in \cite{Cim}, the radial eigenfunction of the fundamental scalar 
mode attains its peak location in correspondence with the minimum of the effective binding potential (\ref{Eq8}) 
of the photon-fluid model.
\bibitem{Notemoi} It is well known that the 
angular velocity of spinning Kerr black holes is characterized by the dimensionless 
upper bound $\Omega_{\text{H}}r_{\text{H}}\leq 1/2$. On the other hand, as discussed in \cite{Cim}, 
there is no fundamental upper bound on the angular velocity of the $(2+1)$-dimensional acoustic black hole (\ref{Eq4}) 
of the photon-fluid model.
\bibitem{Notepwlg} Note that one can assume $\psi(r=r_{\text{H}})\geq0$ for the radial scalar eigenfuntion 
without loss of generality.
\bibitem{Notepdr0} Note that if $d\psi(r)/dr\to0$ and $d^2\psi(r)/dr^2<\infty$ for $r\to r_{\text{H}}$, then one finds 
from Eqs. (\ref{Eq7}) and (\ref{Eq17}) the near-horizon relation $\psi(r\to r_{\text{H}})\to0$. 
This observation together with the asymptotic boundary condition (\ref{Eq11}) imply again that the 
scalar eigenfunction must have an extremum point $r=r_{\text{max}}$ in the exterior region of the acoustic 
black-hole spacetime.
\bibitem{Notemge1} Note that in the last inequality of (\ref{Eq21}) we have used the 
relations $m^2\geq1>1/4>\Delta/4$ for $m\geq1$.
\bibitem{Notevcc} Our goal here is to identify the unique circular trajectory that minimizes the orbital period of test particles
around the central $(2+1)$-dimensional acoustic black hole (\ref{Eq4}) 
as measured by asymptotic observers. We therefore assume
the relation $v/c\to1^-$ for the tangential velocity of the orbiting test particle.
\end{thebibliography}
\end{document}